\documentclass[%
 reprint,
 amsmath,amssymb,
 aps,
]{revtex4-2}

\usepackage{graphicx}% Include figure files
\graphicspath{{./figures/}}
\usepackage{dcolumn}% Align table columns on decimal point
\usepackage{bm}% bold math
\usepackage{hyperref}% add hypertext capabilities
\usepackage{xcolor}

\newcommand*\diff{\mathop{}\!\mathrm{d}}

\begin{document}

\preprint{APS/123-QED}

\title{Exact Coarse Graining Preserves Entropy Production out of Equilibrium}% Force line breaks with \\

\author{Gianluca Teza}
\affiliation{
Department of Physics of Complex Systems, Weizmann Institute of Science,
Rehovot 7610001, Israel
}
\affiliation{
Department of Physics and Astronomy, University of Padova, 
Via Marzolo 8, I-35131 Padova, Italy
}
\author{Attilio L. Stella}%
 \email{attilio.stella@pd.infn.it}
\affiliation{
Department of Physics and Astronomy, University of Padova, 
Via Marzolo 8, I-35131 Padova, Italy
}
\affiliation{
INFN, Sezione di Padova, Via Marzolo 8, I-35131 Padova, Italy
}

\date{\today}% It is always \today, today,

\begin{abstract}
The entropy production rate associated with broken time-reversal symmetry
provides an essential characterization of nanosystems out of equilibrium, from driven colloidal particles to molecular motors.
Limited access to the dynamical states is generally expected to hinder the correct estimation of this observable.
Here we show how memoryless jump processes can be coarse grained exactly preserving its average and fluctuations at stationarity.
This supports univocal applicability of fluctuation theorems for entropy and allows inference of the genuine thermodynamics together with inaccessible process details. 
\end{abstract}

\maketitle
Entropy production out of equilibrium, measured, e.g., from the heat
dissipated by a mesoscopic system into a thermostat, is a key to interpret
experiments involving nanomanipulation or molecular motors~\cite{Liphardt2002,Wang2002,Bustamante2005,
Kolomeisky2007,Lau2007,Astumian2002}.
When dealing with active matter, this production may be the only possible indicator of out of equilibrium conditions~\cite{Martinez2019,Roldan2018}. 
Fluctuations of the rate of produced entropy are expected to obey symmetry properties
that allow us to estimate free energy differences or binding energies at the molecular level~\cite{Hummer2001,Liphardt2002,Mossa2009,Camunas-Soler2017}.
Theorems validating such properties have been proved for specific models~\cite{Evans1993,Gallavotti1995,Kurchan1998,Lebowitz1999}, but do not hold if only part of the states of the mesoscopic system is experimentally accessible. Indeed, applicability of these theorems requires that all slow transitions between mesostates are
detectable, and that transitions inside each mesostate  are very fast~\cite{Mehl2012,Seifert2019}. Under these conditions
entropy production can be recovered from a description without memory in the framework 
of stochastic thermodynamics~\cite{Seifert2012}.

On the other hand, in experiments
where not all mesoscopic details are accessible, like with molecular motors, nonexponential dwelling or residence time distributions have been often measured.\cite{Rief2000,Kolomeisky2000,Bierbaum2013}. These memory effects were identified as revealing features of the underlying chemomechanical transitions, but not put in direct relation to entropy production.

If only partial, coarse-grained information is experimentally available, 
the average entropy production one can record is generally expected to be
lower~\cite{Seifert2019}. Estimates of partial
entropy production pertaining to the accessible parts of the systems, 
or lower bounds for the average full productions, have been
actively studied recently~\cite{Crisanti2012,Kawaguchi2013,Polettini2017,Bisker2017,
Martinez2019,Roldan2018,Gladrow2016}. 
However, 
these results are of limited help for a complete thermodynamic inference~\cite{Alemany2015},  since precise insight on the possible effects of coarse graining on detectable entropy production
is still missing~\cite{Seifert2019}.

The entropy production  of coarse-grained Markov processes governed by Master equations was first studied in cases in which microstates connected by fast transitions rates were lumped into slow evolving  mesostates, implying a sharp timescale separation~\cite{Rahav2007}. So, for large separation the memory effects could be regarded as a small perturbation. More recently, it has been shown~\cite{Esposito2012,Bo2014} that even in the limit of infinite timescale separation the resulting effective slow Markov dynamics generally fails to account for the whole original production.
Thus, establishing the precise effects of coarse graining on the detectable entropy production and finding if there exists one way to coarse grain leaving such production invariant
remains an open fundamental issue, of much relevance for the description of experiments.

In this Letter we present an exact coarse graining by decimation
of Markov jump processes which keeps precise record of 
how the entropy production evolves in time. We also show that its average and higher fluctuation moments remain invariant at stationarity.
Non-Markovian residence time distributions typically recorded in experiments can also be analyzed
in terms of a complementary form of decimation of Markovian trajectories in state space.
The results of this second decimation do not generally allow us to recover the full entropy production, 
but can enable its thermodynamic inference~\cite{Seifert2019,Alemany2015} by reconstruction of the underlying dynamics.

\begin{figure}
\includegraphics[width=0.45\textwidth]{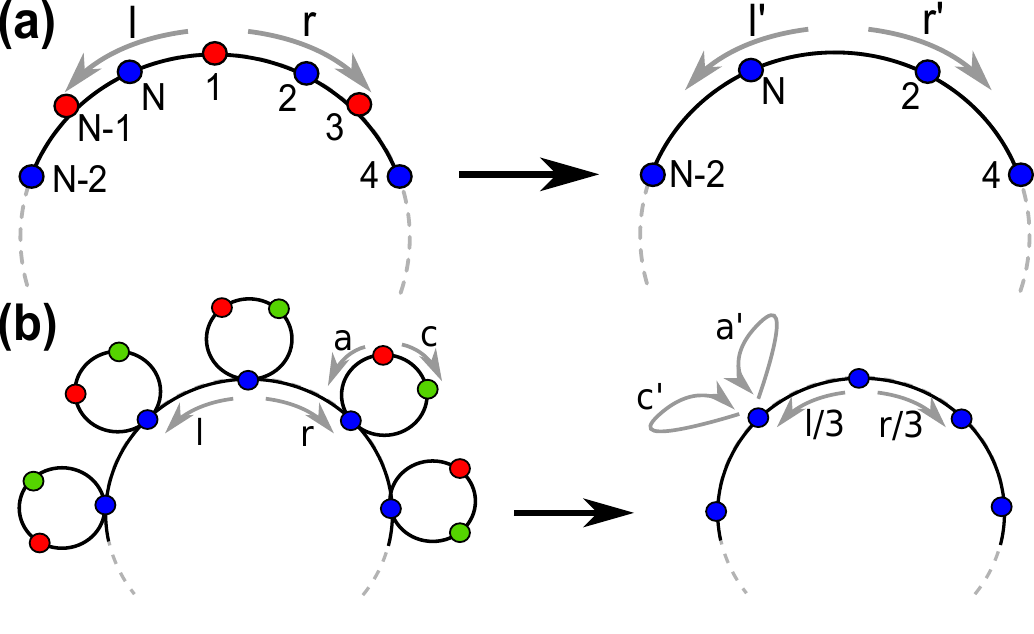}
\caption{\textbf{(a)} Linear network of Eq.~\ref{eq:ME_rl}:
decimation yields the process of Eq.~\ref{eq:ME_rl_CG}, with jumping rates $r'\propto r^2$ and $l'\propto l^2$.
\textbf{(b)} Chain with secondary loops of Eq.~\ref{eq:ME_rlac}: decimation leaves a linear chain in which every site
has additional entropy producing self jumps with rates $a'\propto a^3$ and $c'\propto c^3$.
\label{fig:linear_sketch}}
\end{figure}

We start by considering a Markov process on a linear periodic network with states $i=1,2,\dots,N$ and rates of jump $W_{ji}$ from state $i$ to state $j$. We assume nonzero rates only for
nearest neighbor jumps, and put $W_{ji}=r$ or $W_{ji} =l$,  for all right or left jumps, respectively [Fig.~\ref{fig:linear_sketch}(a)]. If states, e.g., refer to positions of a particle on a lattice with spacing $L$,  $r$ and $l$ can be linked to $L$, to a uniform driving force $f$, and to the thermal bath temperature $T$, by the local detailed balance condition $r/l = e^{f L/k_B T}$~\cite{Katz1983}.
Along a trajectory of the process in which jumps take place at times $t_k$ ($k=1,2,\dots n$;
$0<t_1<t_2<\dots t_n<t$) from states $i_{k-1}$ to states $i_k$ ($i_0$ state at $t=0$) the  entropy produced is ($k_B =1$)~\cite{Lebowitz1999,Seifert2005}
\begin{equation}\label{eq:S_traj}
S=\sum_{k=1}^{n} \log \left[\frac{W_{i_k,i_{k-1}}}{W_{i_{k-1},i_k}}\right]\ ,\end{equation}
with average rate at stationarity compensating exactly the entropy transfered to the medium.

The probability $\overline{P}_i(t)$ for the system to be in state $i$
at time $t$, is marginal of the probability
$P_i(S,t)$ that the system is in state $i$ at time $t$ with an entropy $S$
accumulated along  all possible trajectories. Thus we can write:
\begin{eqnarray}\label{eq:ME_rl}
\nonumber
\left[ r+l+ \partial_t \right] P_{i}(S,t) &&= r P_{i-1}(S-\log (r/l),t) + \\ 
&&+ l P_{i+1}(S+ \log(r/l),t)
\end{eqnarray}
where the shifts in the $S$ arguments on the right-hand side
account for the entropy gains associated with jumps according to Eq.~\ref{eq:S_traj}. 
Summing Eq.~\ref{eq:ME_rl}  over all $S$ values yields the master equation satisfied by
$\overline{P}_i(t)$ with a unique stationary solution reached for $t \to \infty$ from arbitrary initial conditions~\cite{Kampen1992}.
For $t \to \infty$,  $P_i(S,t)$ is consistent with 
a large deviation principle~\cite{Touchette2009} for the entropy production rate $\sigma = S/t$.
Indeed, indicating by $Q(S,t)=\Sigma_i P_i(S,t)$ the probability
of having produced a total entropy $S$ at time $t$, 
the scaled cumulant generating function (SCGF) $\varepsilon$
can be extracted from the function
$G(\lambda,t)=\sum_S e^{\lambda S} Q(S,t) \sim_{t \to \infty} e^{\varepsilon(\lambda,r,l)t}$ \cite{Touchette2009}.
By discrete Laplace-transforming Eq.~\ref{eq:ME_rl} with respect to the entropy $S$ we obtain a first order differential equation for $G$
\begin{equation}
\left[\partial_t +r+l + r e^{\lambda \log r/l}+l e^{\lambda \log l/r} \right]G(\lambda,t)=0
\end{equation}
whose solution, in the long time limit, provides us with the SCGF~\cite{SM}
\begin{equation}
\varepsilon = re^{\lambda\log r/l}+le^{-\lambda\log(r/l)}-(r+l) \ .
\end{equation}
The probability $Q(S,t)$ for $t \to \infty$ concentrates on the value $S= \sigma_0 t$ where $\sigma_0 = \partial \varepsilon / \partial\lambda|_{\lambda=0}=(r-l) \log(r/l)$, while higher order derivatives $\partial^n\varepsilon/\partial\lambda^n|_{\lambda=0}$ give the scaled cumulants of $\sigma$ describing its fluctuations for long times. Since
$\varepsilon$ satisfies the fluctuation theorem~\cite{Lebowitz1999}, i.e.
$\varepsilon(\lambda -1,r,l)=\varepsilon(-\lambda,r,l)$, the probability of $\sigma$ further satisfies $\text{Pr}(S/t=\sigma)/ \text{Pr}(S/t=-\sigma)=e^{\sigma t}$ for $t \to \infty$.

Our coarse graining based on algebraic elimination of probabilities from equations like Eq.~\ref{eq:ME_rl},
can, in principle, be applied for arbitrary choices of the surviving states. The linear character of the network in Eq.~\ref{eq:ME_rl} suggests a choice reducing to a minimum the complexity of the calculations
and preserving the homogeneity of the model.
Indeed, assuming $N>3$ even, one can eliminate from the system in Eq.~\ref{eq:ME_rl} all
odd (or even) states [Fig.~\ref{fig:linear_sketch}(a)].
So,  after Fourier transforming in time Eq.~\ref{eq:ME_rl} one obtains
\begin{eqnarray}
\label{eq:Fourier}\nonumber
\left[ r+l+ i\omega \right] \tilde{P}_{i}(S,\omega) = r \tilde{P}_{i-1}(S-\log (r/l),\omega) + \\
+ l \tilde{P}_{i+1}(S+ \log(r/l),\omega)
\end{eqnarray}
where $\tilde{P}_{i}(S,\omega) = \int_{\mathbb{R}}\ dt e^{i\omega t} P_i(S,t)$, and the odd $\tilde{P_i}$'s
can be algebraically eliminated. 
Thus, upon reverse transforming, the even $P_i$'s satisfy
\begin{eqnarray}\label{eq:ME_rl_CG}
\nonumber
&\left[ \frac{1}{2(r+l)}\partial_t^2 +\partial_t \right] P_{i}(S,t) = \frac{r^2}{2(r+l)} P_{i-2}(S-2 \log(r/l),t) + \\
&+ \frac{l^2}{2(r+l)} P_{i+2}(S+2 \log(r/l),t) - \frac{r^2+l^2}{2(r+l)} P_{i}(S,t) \ .
\end{eqnarray}
Equations \ref{eq:ME_rl_CG} are not consistent with a master equation due to the
$\partial^2_t$ term. The quantity $\sum_S \sum_{i\ \text{even}} P_i(S,t)$
is not conserved, but stabilizes at a value $1/2$ (equivalence of
even and odd states) after a transient time $1/2(r+l)$.
As we show below, in spite of the lack of strict normalization of the $P_i$'s with even $i$, the coarse-grained description provided by Eqs.~\ref{eq:ME_rl_CG} accounts correctly for the entropy production of the original system. Indeed, we can
define $Q'(S,t)=\sum_{i\ \text{even}} P_i(S,t)$ and write a differential equation of second order in time for  $G'(\lambda,t)=\sum_{S} e^{\lambda S} Q'(S,t)$, which becomes the function controlling entropy production in the coarse-grained system:
\begin{eqnarray}\label{eq:PDE_G_rl_CG}
&&\nonumber \left[ \partial_t^2 + 2(r+l)\partial_t + r^2+l^2 \right]G'(\lambda,t) = \\
&&=\left[r^2 e^{2\lambda\log(r/l)} + l^2 e^{-2\lambda\log(r/l)} \right] G'(\lambda,t) \ .
\end{eqnarray}
From the dominant long $t$ behavior $ G' \sim e^{\varepsilon' t}$ of the solution, we argue that the rate $S/t$ still obeys a large deviation principle~\cite{Touchette2009} and find eventually
the SCGF for the coarse-grained entropy production
$\varepsilon'(\lambda,r,l)=\varepsilon \left(\lambda,r,l \right)$~\cite{SM}.
Thus, the SCGF function remains the same as that of the original
process satisfying the fluctuation theorem~\cite{Lebowitz1999}. This is a first instance of our main,
unexpected ~\cite{Seifert2019,Rahav2007,Puglisi2010} result: this decimation leaves
exactly invariant the stationary spectrum
of entropy production fluctuations.

\begin{figure}
\includegraphics[width=0.45\textwidth]{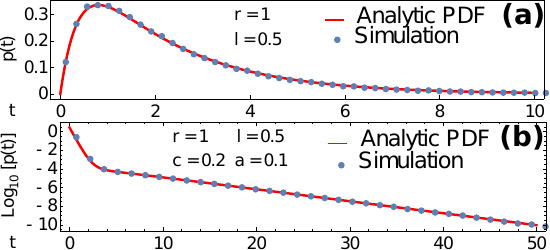}
\caption{
Residence time distributions from decimation of the systems in
%in the coarse-grained systems described by 
Eqs.~\ref{eq:ME_rl} \textbf{(a)} and Eqs.~\ref{eq:ME_rlac} \textbf{(b)}: analytic results (red lines)
%are in agreement with
and data (blue dots) from decimation of trajectories simulated with the Gillespie algorithm~\cite{Gillespie1977,SM}.
\label{fig:res_time}}
\end{figure}

Memory effects in Eqs.~\ref{eq:ME_rl_CG} are represented by the higher-order time derivative terms and each even $\overline{P}_i(t)$ has nonzero contributions only from the trajectories that at time $t$ see state
$i$ occupied, while trajectories visiting odd states at the same 
time do not contribute to any $\overline{P}_i(t)$.
There is a complementary way of decimating the trajectories in which memory reveals in nonexponential probability densities of jump times between surviving states.
In this second form of decimation also trajectories which see odd states occupied at time $t$ contribute to the probability of occupation of the even states.
Indeed, one can ascribe to occupation of an even state $i$ also the time intervals during which state $i-1$ or $i+1$ are visited, up to when state $i+2$ or $i-2$ are first reached~\cite{SM}.
In this way the probability of occupation of the surviving states remains strictly normalized to $1$.
Probability densities of such and related times are often recorded in
experiments~\cite{Rief2000,Kolomeisky2000,Bierbaum2013}, 
and to compute them we introduce a normalized jump rate $w_{ji}=W_{ji}/\sum_k W_{ki}$.
So, for the Markov process, the probability density of the time $t$ of jump from state $i$ to state $j$ after arriving in $i$ at $ t=0$
has the exponential form $w_{ji}e^{-t/\tau}/\tau=\frac{1}{2\pi} \int \diff \omega\  e^{- i\omega t} \tilde{\pi}_{ji}(\omega)$, with $\tau=(r+l)^{-1}$.
The Fourier transform $\tilde{\pi}_{ji}$ allows to express as a series of convolutions the characteristic function of the probability density $p_r(t)$ of jump times from an even state $i$ to the state $i+2$ without visiting $i-2$.
To this purpose we consider the $3\times 3$ matrix $T_{k'k}(\omega) =\tilde{\pi}_{k'k}(\omega)$, restricted to the states $\{i-1,i,i+1\}$.
In this way the characteristic function of
$p_r(t)$ can be written as $\tilde{p}_r(\omega) = \int \diff t\  e^{i \omega t} p_r(t) = \tilde{\pi}_{i+2,i+1}(\omega) (\mathbb{I}+ \mathbf{T}(\omega))^{-1}_{i+1,i}$. This expression sums the contributions of all trajectories performing that jump~\cite{SM}. Analogously one obtains $\tilde{p}_l(\omega)$.
Unlike in Eqs. \ref{eq:ME_rl_CG}, where the $P_i$'s are constructed by recording the
presence of the system exclusively in the even states, one can count the whole times of jump as residence times in state $i$.  With this assumption
the sum $p_r(t) +p_l(t) =p(t)$ becomes equivalent to a non-Markovian (nonexponential) distribution of residence time in a generic even state. This distribution is reported in Fig.~\ref{fig:res_time}a. The process resulting from this interpretation of the jump times
qualifies as semi-Markov with time-direction independence, since $p_r(t) \propto p_l(t)$~\cite{Hughes1995,Esposito2008,Andrieux2008}. The entropy production rate of such models has been determined recently~\cite{Esposito2008,Andrieux2008} and is remarkably consistent with our results for the linear chain~\cite{SM}.
As we show below, in more general cases such consistency does not hold. However, the densities $p_{r,l}(t)$, regarded as empirical data, can be exploited to identify the underlying Markov dynamics.               
Upon matching these densities with analytic results like those just derived, one can determine the rates of the undecimated model and infer the full entropy production with its fluctuations.
In relatively simple cases a successful identification could be guided by physical and analytical insight. At general level establishing conditions under which it would be possible and unique poses a future challenge.

The linear network of Eq.~\ref{eq:ME_rl} presents only one
loop along which
entropy is produced. A fundamental problem left is to study situations in which
coarse graining erases loops producing entropy in the network.
Indeed, detectable entropy production is generally expected to 
lack erased loop contributions and therefore to be lower in such cases~\cite{Puglisi2010,Seifert2019}.

Let us consider the network in Fig.~\ref{fig:linear_sketch}b.
It consists of a main loop of $N$
states (blue), $X_1,X_2,\dots,X_N$,
with right\textbackslash left nearest neighbor jump rates $r$\textbackslash$l$.
A secondary 3-state loop is further attached to
each $X_i$ state along the main loop,
by connecting it to a $Y_i$ (red) and a $Z_i$ (green)
state. Rates $c$ and $a$ apply, respectively, to clockwise
and anticlockwise jumps to nearest neighbors within secondary loops.
The states $Y_i$ and $Z_i$ of each loop are
those we want to decimate.
Before decimation, upon summing over the index $i$ specifying
different $X$, $Y$ and $Z$ states
equations analogous to Eq.~\ref{eq:ME_rl}~\cite{SM}, one gets:
\begin{equation}\label{eq:ME_rlac}
\begin{cases}
\partial_t P_X(S,t) &= r P_X(S-\log\frac{r}{l},t) + l P_X(S-\log\frac{l}{r},t)+ \\
 & + c P_Y(S-\log\frac{c}{a},t) + a P_Z(S-\log\frac{a}{c},t)+\\
&-[r+l+a+c] P_X(S,t) \\
\partial_t P_Y(S,t) &= a P_X(S-\log\frac{a}{c},t) + c P_Z(S-\log\frac{c}{a},t) +\\
&-[a+c]P_Y(S,t) \\
\partial_t P_Z(S,t) &= c P_X(S-\log\frac{c}{a},t) + a P_Z(S-\log\frac{a}{c},t) +\\
&-[a+c]P_Z(S,t)
\end{cases}
\end{equation}
where, e.g., $P_X(S,t)=\Sigma_i P_{X_i}(S,t)$ is the probability that
a trajectory ends at time $t$ in a generic $X$ state with cumulated entropy
$S$. The SCGF $\varepsilon(\lambda,r,l,c,a)$ for
entropy production of this process can be found by considering, for $\alpha=X,Y,Z$, $G_{\alpha}(\lambda,r,l,a,c)=
\sum_S e^{\lambda S} P_{\alpha}(S,t)$ in Eqs.~\ref{eq:ME_rlac} and by
diagonalizing the $3 \times 3$ matrix expressing the time derivative of the
$G_{\alpha}$ vector components.
The dominant eigenvalue determines
$\varepsilon$ \cite{SM} which yields
an average entropy production rate:
\begin{equation}
\sigma_0=\frac{\partial\varepsilon}{\partial\lambda}|_{\lambda=0}   =\frac{r-l}{3} \log \frac{r}{l} + (c-a) \log \frac{c}{a} \ .
\end{equation}
Here the first term is the contribution from jumps on
the main loop,
while the latter is relative to
transitions occurring within the secondary loops. 
Decimation in this case is realized by simply eliminating $P_Y(S,t)$
and $P_Z(S,t)$ from
the system of Eqs.~\ref{eq:ME_rlac} after Fourier transforming in time. Reverse transforming yields a third-order differential
equation in time for $ P_X(S,t)$. Also in this case $\sum_S P_X(S,t)$ is not strictly normalized, but for large $t$ it stabilizes to $1/3$.
Indeed, $X$,$Y$ and $Z$ states have equal total probability $1/3$ at stationarity~\cite{SM}. $P_X(S,t)$ plays here a role analogous to that of $Q'(S,t)$ in the linear chain decimation.
For the function $G'(\lambda,t)=\Sigma_{S}
e^{\lambda S}P_X(S,t)$ of the
coarse-grained network we eventually obtain
\begin{eqnarray}\label{eq:3rd_DE_G}
\left[ \alpha +\beta \partial_t + \gamma \partial_t^2 + \partial_t^3 \right] G'(\lambda,t)=0
\end{eqnarray}
where $\alpha$, $\beta$ and $\gamma$ are functions of
the jump rates and of $\lambda$~\cite{SM}.
In the limit $t\to\infty$ we get $G'\sim e^{t \varepsilon'}$,
where $\varepsilon'$ is the dominant root of the
characteristic equation associated with Eq.~\ref{eq:3rd_DE_G}.
Remarkably, $\varepsilon'$ coincides with the SCGF $\varepsilon$
of the original process~\cite{SM}. So, also in this case
$\varepsilon'=\varepsilon$, maintaining validity
of the fluctuation theorem.
In this example,
the elimination of secondary loops leads to terms in Eq.~\ref{eq:3rd_DE_G}
that represent additional entropy gains originating from self jumps on $X$ states in the coarse-grained dynamics [Fig.~\ref{fig:linear_sketch}(b)].
Such contributions take into account what the removed secondary loops were
producing in the underlying Markov description~\cite{SM}.
Thanks to them, contrary to previous expectations~\cite{Puglisi2010,Seifert2019}, our decimation can keep track of the full entropy production.

Decimation of the process trajectories, carried on along lines similar to those illustrated for the linear chain, allows us to obtain
nonexponential $p_r(t)$ and $p_l(t)$ for the jumps between neighboring $X$ states~\cite{SM}. Also in this case the coarse-grained dynamics can be regarded as semi-Markov with time-direction independence [$p_r(t) \propto p_l(t)$].
In Fig.~\ref{fig:res_time}(b)
we compare the residence time distribution $p(t)=p_r(t)+p_l(t)$ with a histogram based on simulations.

Evaluating the rate of entropy production of this semi-Markov model
according to Refs.~\cite{Esposito2008,Andrieux2008} would account only
for the partial contribution coming 
from direct transitions between surviving $X$ states,  with average $\log(r/l) (r-l)/3<\sigma_0$.
Also procedures setting lower bounds to the average entropy production rate~\cite{Martinez2019} could not improve the estimate of this average, due to the time-direction
independence of the process.
However, even though coarse graining gives
access only to the densities $p_r(t)$ and $p_l(t)$, 
the calculations leading to these
functions still allow us to determine the full entropy production. This can be done by matching the empirical probability densities of jump times
with those of the candidate model for the fine-grained process. Its Markovian dynamics can
be precisely determined together with the full spectrum of entropy production.
The marked difference between the residence time
distributions reported in panel (a) and (b) of Fig.~\ref{fig:res_time}, exemplifies a  useful hint helping
in an attempt to guess the underlying hidden Markov network.

\begin{figure}
\includegraphics[width=0.45\textwidth]{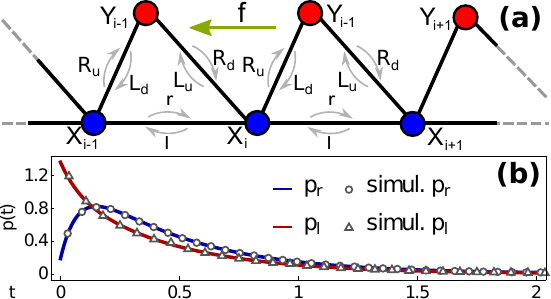}
\caption{\textbf{(a)} Markov jump molecular motor model.  
%Having no access to the active $Y$ states (red), an %observer
%is unable to distinguish between purely mechanical ($r,l%$) and chemomechanical ($R_u R_d , L_u L_d$) %transitions between
%$X$ states (blue). 
Matching $p_r(t)$ and $p_l(t)$ \textbf{(b)}  allows us to recover all rates, full entropy production and ATP consumption.
\label{fig:ATP_motor_sketch}}
\end{figure}

Our coarse graining can be applied, e.g., to molecular motors.
A simplified version of the model in Ref.~\cite{Lau2007} is reported in Fig.~\ref{fig:ATP_motor_sketch}, where the red
dots in the periodic network indicate intermediate hidden $Y$ states allowing ATP  hydrolysis activated
jumps between the blue $X$ states. The positions of the $X$ states are recorded in
experiments.
The transition rates reported in Fig. 3 satisfy local
detailed balance conditions linking them to the load force $f$, the difference in
chemical potentials $\Delta \mu = \mu(ATP) - \mu(ADP) - \mu(P)$, the
temperature $T$ and the spacing $L$ between $X$ positions~\cite{SM}. Elimination from
the equations of the probabilities referring to $Y$ sates leads to
the following system for the probabilities of the $X$ states
\begin{eqnarray}\label{eq:ATP_CG}
\nonumber 
&\left[(r+l+R_d+L_u+R_u+L_d)\partial_t+\partial_t^2\right]P_{X_i}(S,t) = \\
\nonumber
&=r(R_d+L_d)P_{X_{i-1}}\left(S-\log (r/l),t \right)+\ \ \ \ \ \ \ \ \ \  \\
\nonumber
&+l(R_d+L_d)P_{X_{i+1}}\left(S-\log (l/r),t \right)+\ \ \ \ \ \ \ \ \ \ \ \\
\nonumber
&+R_d R_u P_{X_{i-1}}\left(S-\log (R_d R_u/L_u L_d),t \right)+\ \ \ \ \  \\
\nonumber
&+L_u L_d P_{X_{i+1}}\left(S-\log (L_u L_d/R_d R_u),t \right)+\ \ \ \ \ \  \\ 
&-[(r+l)(R_d+L_u)+R_d R_u+L_d L_u ]P_{X_i}(S,t)\ 
\end{eqnarray}
The various terms on the right-hand side account for the entropy contributions due to both direct and ATP activated transitions.
Exact calculations show that this coarse-grained description accounts for the total entropy production also
in this case~\cite{SM}. The process resulting from analysis of the probability densities of jump times is semi-Markov with
time-direction dependence, since  $p_r(t)$ and $p_l(t)$,
now reported in Fig.~\ref{fig:ATP_motor_sketch}(b), are not simply proportional.
The methods of Refs.~\cite{Esposito2008,Andrieux2008} do not allow to exploit this dependence for the computation of the entropy production of such process \cite{Martinez2019}.
On the other hand, if the jump time
probability densities are extracted from some
empirical time series, one can recover the rates of the original Markov model
by matching them with
the results from a decimation of trajectories letting only $X$ states survive.
So, provided the assumption of underlying Markov dynamics is correct, a successful matching allows us to recover the full entropy
produced in the experiment.

The decimation of equations presented above can be performed on networks with arbitrary topology and with states  surviving decimation that are not simply equivalent up to translations. Referring to the simple linear chain case, there is no problem in letting, for example, only a subset of odd or even states survive decimation, even if this breaks the original homogeneity of the transition rates.

An example with inhomogeneity allowing us to make contact with the timescale separation situations considered
in Refs.~\cite{Rahav2007,Esposito2012,Bo2014}, is that of a linear chain where
nearest neighbor links having fast rates, $R$ and $L$,  alternate with 
links having slow rates, $r$ and $l$. In the decimated equations only one state for each pair of two fast connected states survives to represent the whole cluster, and the coarse-grained dynamics again preserves entropy production for arbitrary timescale separations. This example is treated in detail in Ref.~\cite{SM}. One can also keep simultaneous record of other currents besides the entropy production rate~\cite{Teza2020}.

Summarizing, we showed that, in the context of Markov jump
processes, an exact coarse graining, taking into account
memory effects, and guaranteeing
invariance of average and fluctuations of
the full entropy production  at stationarity, is possible.
While an extension of the type of exact results
presented here to other models is an open program,
our findings suggest that future investigations in
this field, both theoretical and experimental, should focus
on the memory effects associated with coarse graining,
which revealed essential to guarantee entropy production invariance~\cite{Crisanti2012}.
These effects, when regarded as a result of our decimation of trajectories, can be also crucial for a thermodynamic inference strategy.
Indeed, after detecting non-Markovian probability densities
for the times of jump in an experimentally studied process,
one can try to match them with those of a guessed underlying Markov process.
If successful, this  enables
to determine the correct full entropy production,
while uncovering hidden states 
and mechanisms.
An assumption at the basis of this strategy
is that  Markov jump dynamics is 
adequate to describe what underlies the coarse-grained level.
Such dynamics should, in principle, result from coarse-graining of a more
microscopic description, and the choice of rates may
reveal crucial for the very possibility of describing certain
phenomena~\cite{Teza2019}.

\begin{acknowledgments}
We acknowledge Carlo Vanderzande, Marco Baiesi and Stefano Iubini for collaboration on related subjects.
We thank David Mukamel and Oren Raz for discussions. G.T. is supported by a research grant from the Center of Scientific Excellence at the Weizmann Institute of Science and by the grant of Simons Foundation.
\end{acknowledgments}

\bibliography{references}% Produces the bibliography via BibTeX.

\end{document}